# Controlling the bursting size in the two-dimensional Rulkov model

Jennifer López,[1] Mattia Coccolo,[1] Rubén Capeáns,[1] and Miguel A.F. Sanjuán[1,2]

[1]*Nonlinear Dynamics, Chaos and Complex Systems Group,*
*Departamento de Física, Universidad Rey Juan Carlos,*
*Tulipán s/n, 28933 Móstoles, Madrid, Spain*
[2]*Department of Applied Informatics, Kaunas University of Technology,*
*Studentu 50-415, Kaunas LT-51368, Lithuania*

(Dated: December 21, 2022)

## Abstract

We propose to control the orbits of the two-dimensional Rulkov model affected by bounded noise. For the correct parameter choice the phase space presents two chaotic regions separated by a transient chaotic region in between. One of the chaotic regions is the responsible to give birth to the neuronal bursting regime. Normally, an orbit in this chaotic region cannot pass through the transient chaotic one and reach the other chaotic region. As a consequence the burstings are short in time. Here, we propose a control technique to connect both chaotic regions and allow the neuron to exhibit very long burstings. This control method defines a region $Q$ covering the transient chaotic region where it is possible to find an advantageous set $S \subset Q$ through which the orbits can be driven with a minimal control. In addition we show how the set $S$ changes depending on the noise intensity affecting the map, and how the set $S$ can be used in different scenarios of control.

## I.    INTRODUCTION

Neurons are complex entities that form a highly structured network. In recent decades, it has been of interest to create mathematical models that mimic their behavior. Due to their high complexity, these models have to be approximated and simplified while retaining only certain functionalities of the neurons. The entire structure of the neuron is usually replaced by a system with a voltage membrane and a connection topology.

Initially, continuous models such as the models of FitzHugh-Nagumo [1], Hindmarsh-Rose [2], and Hodgkin-Huxley [3] have been used profusely. Recently, discrete models have started to arouse interest for the simulation of neurons [4]. They are easier models to solve (they avoid the integration of ordinary differential equations), very versatile when creating neural networks and, in addition, they offer results very attractive in terms of dynamical behavior. In this context, it is common to use two-dimensional maps of the slow-fast type variables (fast-slow system). The most relevant are [4]: the Izhikevich's discrete model, Courbage's model, Chialvo's model, and the Rulkov model. The latter presents three variants of the model: non-chaotic, supercritical, and chaotic.

The chaotic Rulkov neuron map [5–11] is used in this work, because it is a simple model that can exhibit the basic regimes of neuronal activity, as rest, bursts, spikes, with the last two allowing periodic and chaotic dynamics.In particular, we focus our attention in the regime where the Rulkov map exhibits chaotic cycles. These cycles alternate periods with high activity of the neuron exhibiting fast chaotic oscillations (bursting), together with periods of low activity without chaotic motion. Typically the bursting size is short because these chaotic oscillations takes place in a narrow chaotic region of the phase space delimited by the presence of an unstable manifold. Once the chaotic orbit touches this unstable manifold the bursting rapidly extinguished giving way to a low activity period.

However, we found that it is possible to greatly increase the bursting size of the neuron exploiting the fact that there is a second chaotic region in the phase space. This second chaotic region is separated from the main chaotic region, where the bursting of the neuron takes place. In this work we explore the possibility to built a path in the phase space connecting both chaotic regions, allowing the chaotic orbits to pass from one to another, and therefore extending the size of the burstings. To do that, we present a control method inspired in our previous work of partial control [12–18] that also takes into account noise affecting the map, as in all real systems. This method defines a region $Q$ in the phase space located between both chaotic regions. Through a recursive algorithm it can be computed a special subset called $S \subset Q$ through which the orbits can be controlled to go from one chaotic region to the other, minimizing the need of control. Furthermore we will see that this method adapts to the intensity of noise affecting the map. Different intensities result
in different sets $S$.

Although this control method resembles the partial control method and they share similarities in the steps to apply it, we want to emphasize that there is a substantial difference in the control goal. While partial control is designed to keep the orbits forever in the region $Q$ of the phase space, this control method is designed to steer the orbits through the region $Q$, allowing the orbits to enter or abandon it via a portion of its boundaries, previously set by the controller, as it is shown schematically in Fig. 1. This control method is specially
indicated to connect different regions of phase space that otherwise would be isolated .

The manuscript is organized as follows. In Sec. II, we introduce the model system. In Sec. III, we describe the control technique. Then, in Secs. IV and V, we apply the control technique to the system in different scenarios, where we also show results for different noise intensities to illustrate how the set $S$ changes. In Sec VI we discuss the results when one of the variables is not controlled or affected by the noise. In Sec. VII we discuss how to generalize the method to other systems. Finally, in Sec. VIII we summarize the main results of the paper.

## II.     THE TWO-DIMENSIONAL RULKOV MAP

The chaotic Rulkov model [5–8] is a two-dimensional map that achieves to exhibit the basic regimes of neuronal activity with a simple model.
The equations of the system are:

$$x_{n+1} = \frac{\alpha}{(1+x_n^2)} + y_n$$
$$y_{n+1} = y_n - \sigma x_n - \beta, \tag{1}$$

being *x* the voltage of the neuron membrane (taking the role of the fast variable), and *y* the ion concentration (representing the slow variable) where $\alpha, \sigma$ and $\beta$ are the system parameters. Here we are interested in the regime where the system exhibits chaotic oscillations

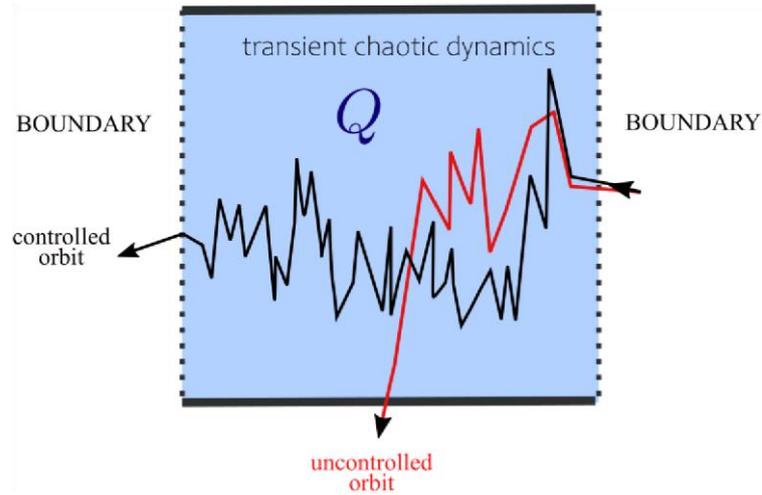

Figure 1: **Control goal.** $Q$ is a region in the phase space previously defined by the controller. In absence of control, an orbit (red orbit) enters in $Q$ through the right boundary (right dashed line) but never reaches the left boundary (left dashed line), since it abandons $Q$ through the bottom boundary. With the suitable application of control it is possible to sustain the orbit (black orbit) in $Q$ until it reaches the right boundary. In this way, the region $Q$ acts as a pathway for the controlled orbits, connecting different parts of the phase space.

[19], so we fix the values $\alpha = 4.1$, $\sigma = \beta = 0.001$, following Rulkov's original article [5].

The behaviour of the orbits in the phase space is shown in Fig. 2. At first glance, the figure looks like a bifurcation diagram but that is not the case. In Eq. 1, both variables $x$ and $y$ are changing. However the change of the $y$ variable is so slow in comparison with the variable $x$, that it behaves almost as a parameter, and that is why the orbits in the phase space, shown in Fig. 2, resembles a bifurcation diagram.

To build the Fig. 2, we took a grid of initial conditions in the rectangle $(y,x) \in [-4.5,-2.5] \times [-5,2]$ and for each initial condition we simulate 100 iterations of the corresponding orbit. We remove the first 99 iterations and we display the iteration 100. In this way, we can synthesize and obtain qualitative information about the behavior of the orbits in the phase space. It is possible to appreciate the diversity in the dynamics that offers this system. We indicate in the figure four important points $(y_1, y_2, y_3, y_4)$. At points $y_1$ and $y_4$, the stable an unstable manifold (displayed in green) intersects, and therefore the orbits changes their stability. Between the points $y_2$ and $y_3$, transient chaotic dynamics takes place. Orbits in this region quickly decay below the unstable manifold and reach the bottom stable

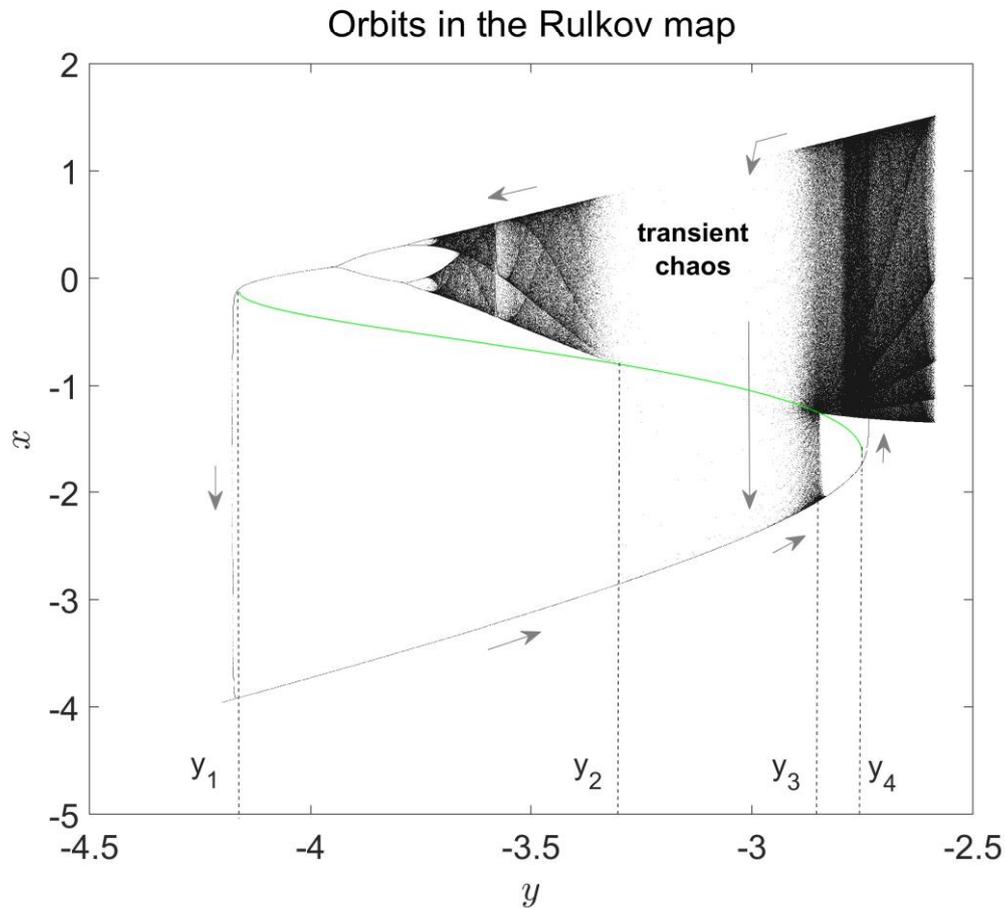

Figure 2: **Behavior of the orbits in the Rulkov map.** Here, we take an uniform grid of 1623 × 2739 initial conditions in the rectangle $(y,x) \in [-4.5,-2.5] \times [-5,2]$. For each initial condition we compute 100 iterations of the corresponding orbit, computed with Eq.1. The figure represents the positions of the orbits in the iteration 100*th*, the previous 99 iterations have been removed to visualize the qualitative behavior of the orbits in each part of the phase space. Notice that this is not a bifurcation diagram since the *y* value also change in every iteration of the orbit. However as *y* is the slow variable, it behaves almost like a parameter and that is the reason the figure resembles a bifurcation diagram. The small arrows displayed, indicate the average motion of the orbits in each region of the phase space. At the points $y_1$ and $y_4$ the stable and unstable manifolds (draw in green) meet and orbits at these points change their stability. The points $y_2$ and $y_3$ delimit the region where the map exhibits transient chaos. Orbits in this region quickly decay below the unstable manifold where they return

to $y_4$. Sooner or later, the orbits of all initial conditions in the rectangle eventually end in the chaotic cycle around $y_3$ and $y_4$.

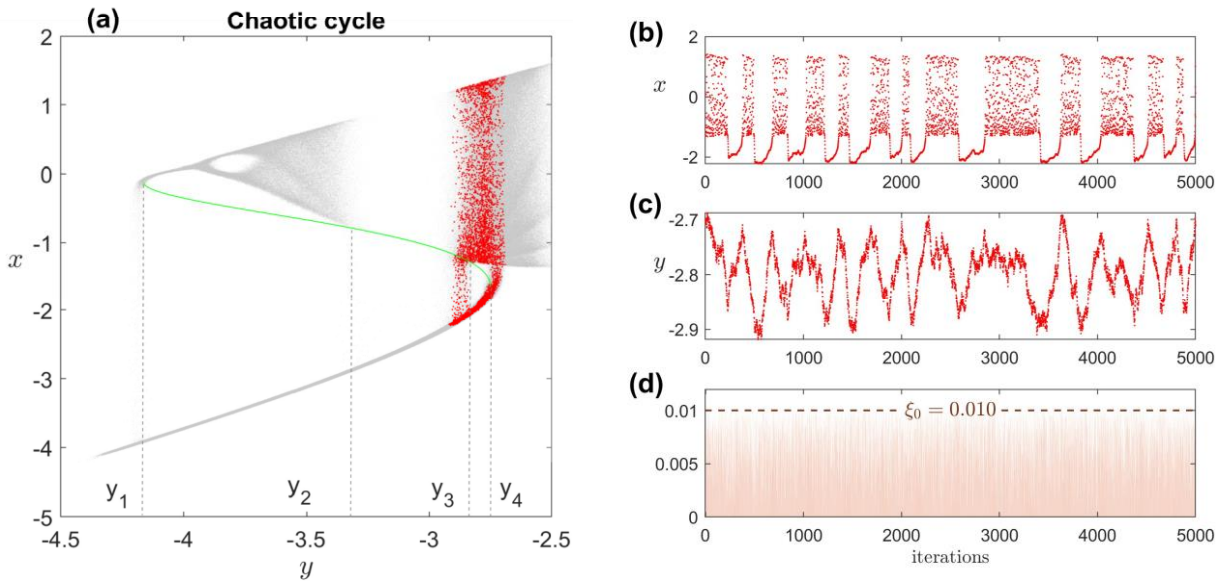

Figure 3: **Chaotic cycle affected by disturbances.** (a) The background orbits shown in grey have been computed in the same way as the Fig. 2 but instead, these orbits correspond with Eq.2 where an upper bound of disturbance $\xi_0 = 0.010$ has been taken. Eventually all these orbits converges to the chaotic cycle (red dots) that remains confined around $y_3$ and $y_4$. The orbit displayed consists of 5000 iterations and the corresponding $x$ (fast variable) and $y$ (slow variable) time series of the red orbit are shown in (b) and (c) respectively. In (d) the disturbances $|\xi_n| \leq \xi_0$ affecting the orbit.

manifold.

To explain how the orbits behave in the phase space (see Fig. 2), let's take an orbit starting in some point on the left chaotic region ($y < y_2$). Here the orbit quickly oscillates in the vertical axis ($x$-axis) while it slowly moves to the left ($y$-axis) towards the periodic region where, eventually, it reaches the point ($y = y_1$). At this point, the orbits touch the unstable manifold and fall to the stable manifold at the bottom. Here the orbits starts to move to the right along the stable manifold until it reaches the value ($y = y_4$), where the orbits meet again the unstable manifold and jumps to the right chaotic region at the top. In this region the orbit starts to oscillate chaotically, while it slowly moves to the left. Finally the orbit reaches the crisis point ($y = y_3$), and it falls again in the bottom stable manifold, repeating forever the chaotic cycle around the values $y_3$ and $y_4$.

To model a more real behaviour of the neuron, we consider that Eqs. 1 are affected by some additive bounded noise, that we call disturbance. In the literature, we can find authors that consider the disturbance affecting only the fast variable *x* [6, 8]. Others consider the disturbance affecting the slow variable *y* [20] and others consider a disturbance affecting both variables [7, 21, 22]. In this work, we consider this last case for being the most general, and at the end of the paper we analyze the particular cases of the disturbance affecting only one variable. The Rulkov map affected by a disturbance is given by:

$$x_{n+1} = \frac{\alpha}{(1+x_n^2)} + y_n + \xi_n^x$$
$$y_{n+1} = y_n - \sigma x_n - \beta + \xi_n^y, \qquad (2)$$

where $\xi_n^x$ and $\xi_n^y$ are the disturbances on each variable. Physically, the disturbance in *x* can represent, for example, the synaptic input noise in the neuron membrane voltage, while the disturbance in *y* models ion-concentration fluctuations, which may be either from outside the cell or from inside [23]. The only condition that we impose is that the disturbance is bounded as $\sqrt{(\xi_n^x)^2 + (\xi_n^y)^2} \leq \xi_0$. In this way, we are confident that it does not become too large compared to the orbits.

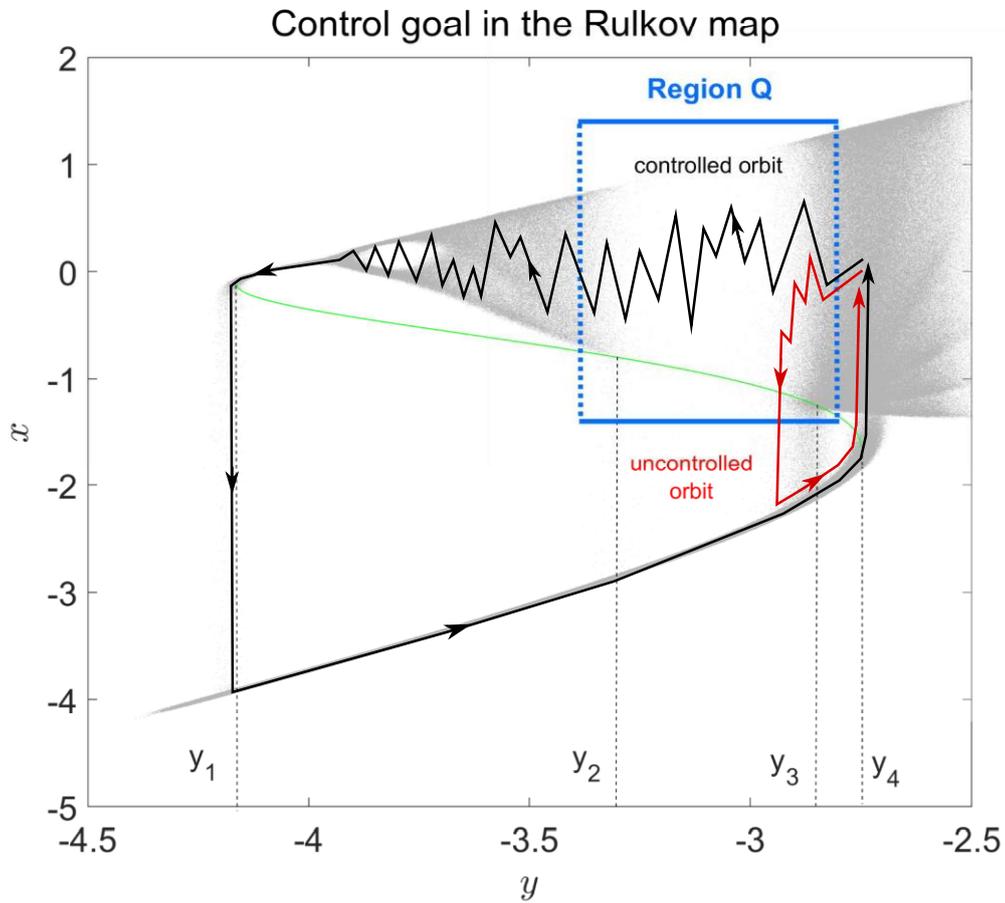

Figure 4: **Scheme of the control goal in the phase space of the Rulkov map**. The background orbits shown in grey are the same as displayed in Fig. 3. They will also be displayed in other figures as a background reference to help the visualization of the uncontrolled and controlled orbits in the phase space. In red color an uncontrolled orbit and in black a controlled one. Both are draw schematically. The control is applied in $Q$ to sustain the transient chaotic orbit and allow it to complete a long bursting. Note that the region $Q$ is taken wider than the interval between $y_3$ and $y_2$ interval. This is because, the orbits affected by disturbances can touch the unstable manifold (green line) before $y_3$ or after $y_2$ and fall to the stable manifold at the bottom. Also, the right and left sides of $Q$ are defined as open boundaries (dashed blue lines) to allow the orbit to enter and escape from $Q$.

The behavior of the noisy orbits in the Rulkov map given by Eq. 2 is displayed in Fig. 3. In grey we display many orbits in the phase space taking a grid of initial conditions in the rectangle $(y,x) \in [-4.5,-2.5] \times [-5,2]$. The grey orbits have been computed in the same way as the orbits displayed the Fig. 2 but using instead the Eq. 2 with an upper disturbance bound $\xi_0 = 0.010$. Eventually all these orbits end in the chaotic cycle displayed by the red dots. In the same figure, we also display the *x* and *y* time series corresponding to the chaotic cycle and the disturbances $|\xi_n| \leq \xi_0 = 0.010$ affecting it. Notice that due to the disturbances, the orbit can touch the unstable manifold before reaching the points $y_3$ and $y_4$, respectively, and therefore the bursting sizes are more irregular in comparison with the deterministic case ($\xi_0 = 0$), but yet short in time.

In this scenario, we propose a control technique to increase the bursting size taking advantage of the presence of the transient chaotic region between $y_2$ and $y_3$ and the left chaotic region. Normally, the chaotic cycles trapped in the right chaotic region could never reach the left chaotic region. However, with a suitable application of control it is possible to sustain the chaotic orbits in the transient chaotic region and allow them to reach the left chaotic region, extending the bursting size of the neuron as it is schematically draw in Fig. 4.

### III. CONTROL SCHEME

As shown in the Fig. 4, when an orbit enters in the transient chaotic region, approximately $y_2 < y < y_3$, after a short transient, it touches the unstable manifold (green line) and fall towards the stable manifold at the bottom. To avoid this escape, we will apply control in the region $Q$ defined as the rectangle $(y,x) \in [-3.42,-2.78] \times [-1.82,1.92]$. This region is $y$-wide enough to contain the interval $y_2 < y < y_3$, and $x$-wide enough to allow the chaotic oscillation of the fast variable $x$ and therefore, preserving the dynamical behavior of the burstings in this region.

In this control scheme, we consider the general case where the control is applied on both variables. At the end of the paper we particularize to the case where the control is only applied on only one variable. The Rulkov map with control in both variables is given by:

$$x_{n+1} = \frac{\alpha}{(1+x_n^2)} + y_n + \xi_n^x + u_n^x \quad (3)$$
$$y_{n+1} = y_n - \sigma x_n - \beta + \xi_n^y + u_n^y$$

where the disturbance is bounded so that $\sqrt{(\xi_n^x)^2 + (\xi_n^y)^2} \leq \xi_0$, and the control applied is also considered bounded so that $\sqrt{(u_n^x)^2 + (u_n^y)^2} \leq u_0$. To simplify the notation, we define the state vector $q_n = (x_n, y_n)$, the disturbance vector $\xi_n = (\xi_n^x, \xi_n^y)$ and the control vector $u_n = (u_n^x, u_n^y)$ so that the map given by Eq. 3 can be written as:

$$q_{n+1} = f(q_n) + \xi_n + u_n, \quad (4)$$

with $|\xi_n| \leq \xi_0$ and $|u_n| \leq u_0$. This upper control bound $u_0$ is specified by the controller but we have to take into account that not any $u_0$ value is possible. There is a minimum value $u_0^{min}$ for which exist points in $Q$ that are controllable. These points constitute a subset of $Q$ that we name the set $S$. Higher values of $u_0 > u_0^{min}$ result in a larger set $S$.

The computation of the set $S \subset Q$ can be realized through a recursive algorithm. Beginning from the set $Q_0 = Q$, the points $q_n \in Q$ for which the image $f(q_n) + \xi_n + u_n$ can not be put it back again in $Q$ with $|u_n| \leq u_0$, are removed. Notice that, for every point $q_n$, all possible disturbances $|\xi_n| \leq \xi_0$ must be evaluated. If for any of these disturbances, the point can not be controlled, then the point $q_n$ is removed from $Q_0$. There is only one exception to this rule. The points $q_n \in Q_0$ for which the image $f(q_n) + \xi_n$ abandon $Q_0$ through the right or left boundary are not removed. This exception is required since we want the controlled orbits to pass across the region $Q$ and leave it through the right or left boundary. In that sense we want that $Q$ actuates like a bridge connecting the right ($y > y_3$) and the left ($y < y_2$) chaotic sides of the phase space and preventing that the orbit escapes through bottom ($x = −1.82$) boundary of $Q$.

After removing all the uncontrollable points $q_n \in Q_0$ in the first iteration of the algorithm, the surviving points constitutes a new subset $Q_1 \subset Q_0$. The second iteration of the algorithm consists on repeating the process described before, but with the subset $Q_1$ instead of $Q_0$. After that we obtain the subset $Q_2 \subset Q_1 \subset Q_0$. In the next steps, the algorithm is repeated until it converges, that is when $Q_{i+1} = Q_i$. This final set will be $S$. This set guarantees that any point $q_n \in S$ can be controlled in $S$ applying every iteration a control $|u_n| \leq u_0$, unless the orbit abandons $Q$ across the right or left boundaries. In that instant the applications of control is stopped.

The computation of the set $S$ as described above, can be greatly speeded up with the following algorithm based on morphological transformations of $Q$. Given the initial region $Q_0 = Q$ and the upper bounds $\xi_0$ and $u_0$, the *ith* step of the algorithm is summarized in Fig. 5.

Notice that if the value $u_0$ selected is too small, the final set $S$ will be the empty set (no points in $Q$ are controllable with such a small control) and therefore we have to select a bigger value $u_0$. As controllers, we want to keep the amount of control as low as possible, so it is reasonable to try to find out the minimum $u_0$, named $u_0^{min}$, for which the set $S$ exists. To do that, we compute the set $S$ several times, taking each time a value $u_0$ closer to the $u_0^{min}$.

That is, for a given value $u_0$, if the set $S$ exists, then we compute it again with a smaller value $u_0$. If the set $S$ is empty, we compute it again with a bigger value $u_0$. In that way we can approximately find the $u_0^{min}$. All the sets $S$ shown in this work were computed with a value $u_0$ very close to the $u_0^{min}$ so the sets $S$ are minimal. Any other set computed with a bigger value $u_0$ will contain the minimal set $S$.

In order to compute an example, we choose the upper disturbance bound affecting the map to be $\xi_0 = 0.010$. For this value we found that the minimum control bound for which the set $S$ exists is approximately $u_0 = 0.008$. After applying the recursive algorithm, we

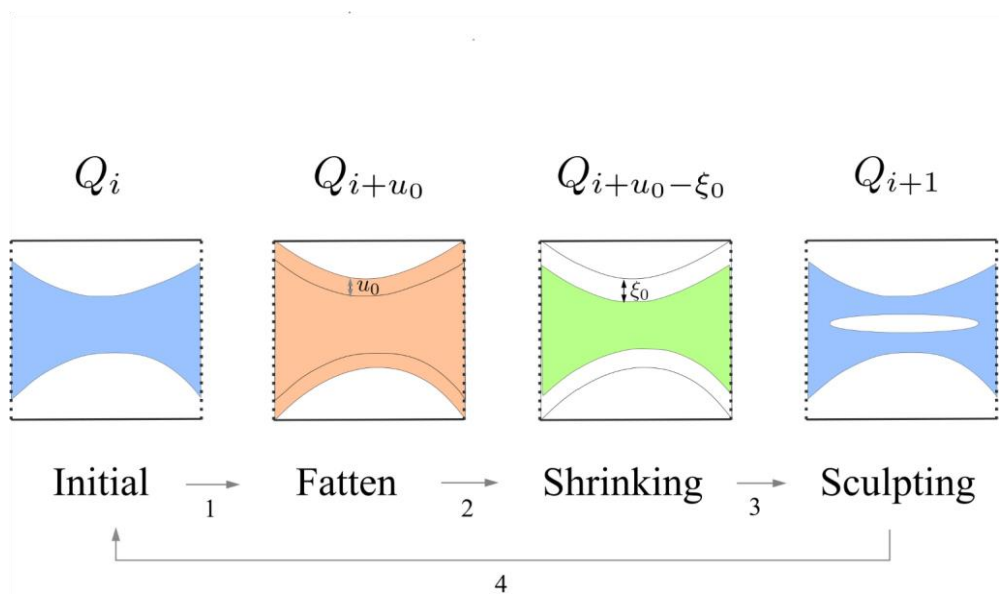

Figure 5: **Recursive algorithm to compute the set $S \subset Q$. Beginning with $Q_0 = Q$.**

**Step 1.** Fatten the set $Q_i$ by $u_0$ except the right and left boundaries, obtaining the set denoted by $(Q_i + u_0)$.

**Step 2.** Shrink the set $(Q_i + u_0)$ by $\xi_0$ except the right and left boundaries, obtaining the set denoted by $(Q_i + u_0 - \xi_0)$.

**Step 3.** Let $Q_{i+1}$ be the points $q \in Q_i$, for which $f(q)$ fall inside the set denoted $(Q_i + u_0 - \xi_0)$, or the points $q \in Q_i$ for which $f(q)$ abandon $Q$ through the right or left boundaries.

**Step 4.** Return to step 1, unless $Q_{i+1} = Q_i$. We call this final region, the set $S$.

obtain the set $S$ shown in Fig. 6, where we also display the 29 iterations that the algorithm takes to converge, from $Q_0$ to $Q_{29}$. In the following subsections we describe three different scenarios that we consider of interest, where the orbit is controlled in $S$ to extend the chaotic bursting of the neuron.

## IV. CONTROL IMPLEMENTATION USING THE SET $S$

In this section we use the set $S$ computed in the previous section to control the orbits. Although the set $S$ was computed to sustain the chaotic orbit through all the region $Q$, we will show that the set *S* can be also used to control the bursting size in *Q*. Here we distinguish the following three scenarios of control implementation.

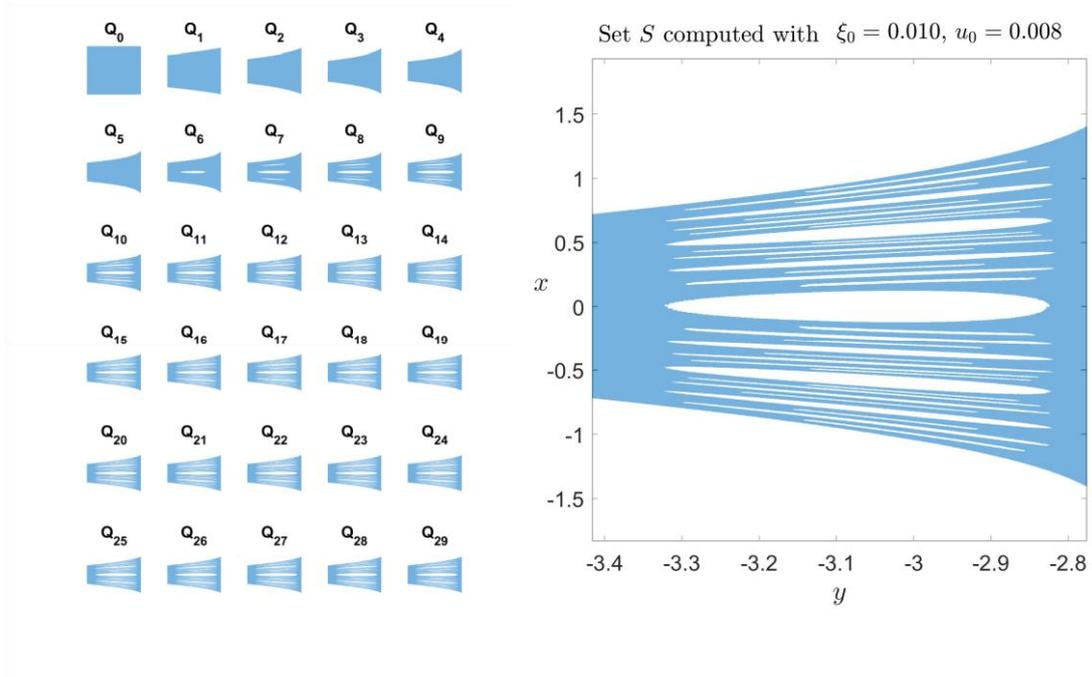

Figure 6: **Computation of the set $S$ with $\xi_0$ = 0.010 and $u_0$ = 0.008**. The region $Q$ is taken as the rectangle $(y,x) \in [-3.42,-2.78] \times [-1.82,1.92]$. The right and left sides of $Q$ are open boundaries. The grid resolution taken in $Q$ is 1000 × 1000 points. The computation of the set $S$, starting from $Q_0$, takes 29 iterations to converge (see the left small figures). In this case the set $S$ corresponds to $Q_{29}$ shown in bigger size on the right.

### A. Control through all the region *Q* (the long bursting)

In this scenario, we control the orbit in *Q* to allow them to achieve the left side of *Q*. All we have to do when the orbit enters in *S* is to apply every iteration of the map $q_{n+1} = f(q_n) + \xi_n + u_n$ the corresponding control $|u_n| \leq u_0 = 0.008$ to keep the orbit inside *S* until it escapes through the left boundary.

In Fig. 7, we show the result of controlling the orbit through all the region *Q*. As shown in Fig. 7(a), the bursting size is greatly increased as can be seen if we compare the *x*-time series shown in Fig. 7(b) and the *x*-time series corresponding to the uncontrolled orbit shown before in Fig. 3(b). In Figs. 7(c) and 7(d), we also show the $y$-time series and the disturbance and the control affecting the 5000 iterations of the orbit. Notice that in this scenario, the chaotic oscillations (bursting) comes with a final periodic oscillation, so that in the high activity period of the neuron, both behaviors are present.

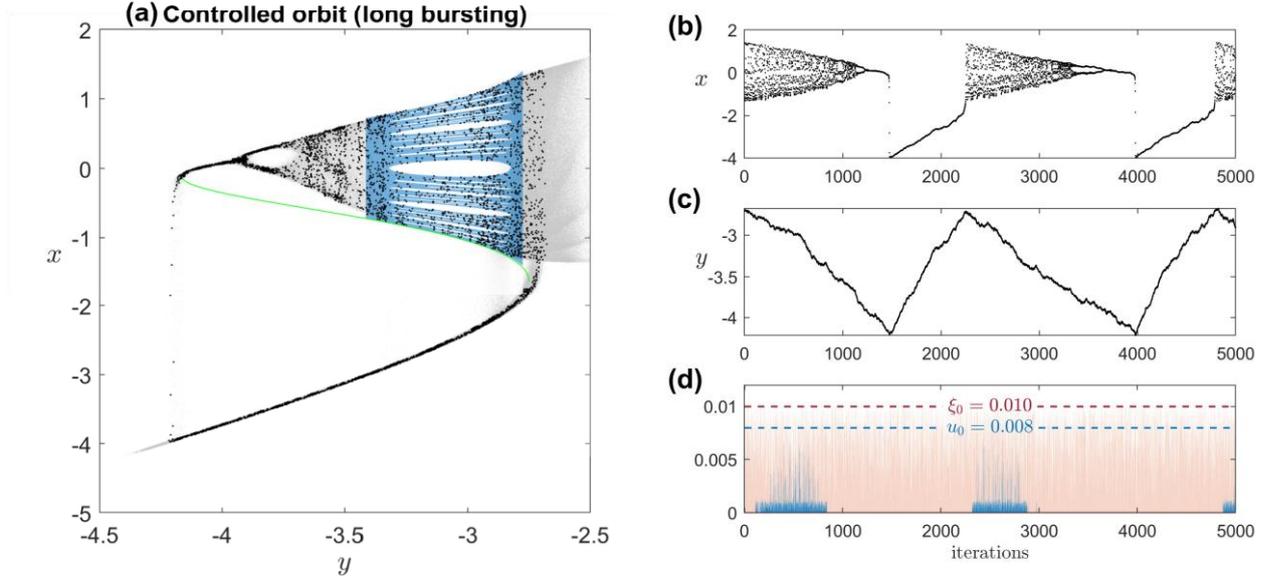

Figure 7: **Long bursting control.** (a) In blue the set $S$ computed for $\xi_0$ = 0.010 and $u_0$ = 0.008. In this set, the control is applied to the orbit (black dots) allowing it to reach the left chaotic region and thus completing a long bursting. (b) The $x$-time series of the controlled orbit. (c) The $y$-time series of the controlled orbit. (d) The disturbances $|\xi_n|$ (orange bars) affecting the orbit and the controls $|u_n|$ (blue bars) applied during the 5000 iterations of the orbit. This control never exceeds the value $u_0$ = 0.008.

### B. Control until a specific $y$ value in the set $S$ (the $y$-stop).

In this subsection and the next one, we show how we can use the set $S$ to control the bursting size of the neuron. In particular, here we analyze the possibility of stop the bursting when the orbit reaches a certain $y$ value inside $Q$.

To compute an example, we choose the limiting value $y_{stop}$ = −3.1 (which is inside the set $Q$). Once the controlled orbit reaches this value, we just cease the application of control. Next, after a short chaotic transient, the orbit naturally escapes from $Q$ through the bottom boundary and the bursting stops. Then, the orbit returns through the stable manifold to initiate the next bursting cycle.

This simple method of stopping the bursting works well in this system. However, depending on the disturbance affecting the transient chaotic orbits, they can take different times to escape from $Q$. A good strategy to reduce this time is, when the orbit reaches the value $y_{stop} = -3.1$, to continue applying a control $|u_n| \leq u_0$, but now with the aim of pushing the orbit as far as possible from the set $S$. This approach significantly reduces the escape time of the orbit.

The result of this control is shown in Fig. 8. In Figs. 8(a) and 8(b), it can be appreciated that the bursting is abruptly stopped when the controlled orbit reaches the value $y = -3.1$. Then, the $y$ variable starts to grow again, see Fig. 8(c). Note that in this case, the control is only applied in $Q$, first to keep the orbit in the set $S$, and then to accelerate the escape from it. This is clearly shown in Fig. 8(d) where the disturbance and the control applied to the orbit are also displayed.

In this scenario of control, it is important to stress out that the bursting cycles have different size (see Fig. 8(a)). This is mainly due to the fact that the slow variable $y$ that leads the cycle is affected by disturbances, just as the $x$ variable, and therefore every bursting can take a different number of iterations to reach the stopping value $y = -3.1$. The higher the upper disturbance bound, the more different bursting size we found. To achieve more similar cycles we propose an alternative control strategy in the next subsection.

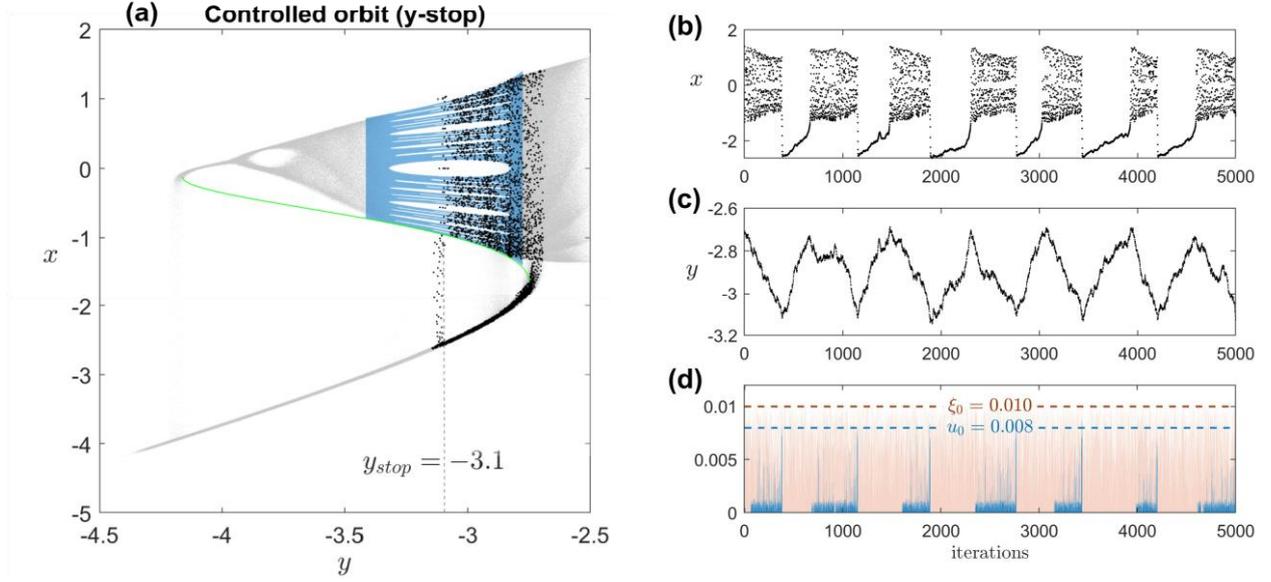

Figure 8: **Control until a specific *y* value in set *S*.** (a) In blue the set *S* computed for $\xi_0$ = 0.010 and $u_0$ = 0.008 . In this set, the control is applied to sustain the orbit (black dots) until it reaches the value $y = -3.1$. Then the escape of the orbit from *S* is forced. (b) The *x*-time series of the controlled orbit. (c) The *y*-time series of the controlled orbit. (d) The disturbances $|\xi_n|$ (orange bars) affecting the orbit and the controls $|u_n|$ (blue bars) applied during the 5000 iterations of the orbit. Note that the control is only applied inside *Q* and it never exceeds the upper bound value $u_0$ = 0.008.

### C. Control to obtain cycles with a similar size

What we pursue here, is to obtain bursting cycles with approximately the same size. To do that, we stop the bursting regime when it reaches certain number of iterations. The only requirement is that the orbit has to be in *Q*. Here, as an example, we choose to stop the bursting when the bursting reaches 600 iterations. However this condition is not enough to achieve similar burstings size because the *y*-variable is affected by the disturbance in all the chaotic cycle, (i.e., the bursting period in *Q* and in the low activity period outside *Q*), and therefore we need to control the *y*-variable during all the cycle.

To do that, we assume that we know the behaviour of the map without disturbances. Taking into account that this deterministic map produces chaotic cycles with similar sizes, we can use the *y*-variable of this deterministic map (we call it $y^*$), to lead the *y*-variable of our map affected by disturbances. In that way we can achieve cycles with similar size.

Combining the above control of the variable *y* along all the cycle, and the control of both variables *x* and *y* in the region *Q*, and taking into account the constraint $|u_n| \leq u_0$) in each iteration, we propose the following full scheme of control:

- For a given point $q_n$ of the orbit, if we want that the image $q_{n+1} = f(q_n) + \xi_n + u_n$ maps in *S*, among all the possible points $q_{n+1} \in S$ (reachable with $|u_n| \leq u_0$) we choose the point for which $|y - y^*|$ is smaller.

- For a given point $q_n$ of the orbit, if we want that the image $q_{n+1} = f(q_n) + \xi_n + u_n$ maps outside *S*, among all the possible points $q_{n+1} \in/ S$ (reachable with $|u_n| \leq u_0$) we choose the point for which $|y - y^*|$ is smaller.

The result of this control scheme is shown in Fig. 9(a). This figure is very similar to Fig. 8(a), nevertheless it should be noticed that in this case, the bursting is stopped when the bursting duration reaches 600 iterations, instead of stopping when the orbit reaches the value $y = -3.1$. Furthermore, as a result of controlling the *y*-variable all the time, the resulting cycles have approximately the same size as shown in Fig. 9(b). See also that the y-series of the controlled orbit, Fig. 9(c), is much more smooth than the *y*-series presented in Fig. 8(c). The counterpart of this control scheme is that now, the amount of control used is larger (see Fig. 9(d)) but always below the upper control bound $u_0 = 0.008$.

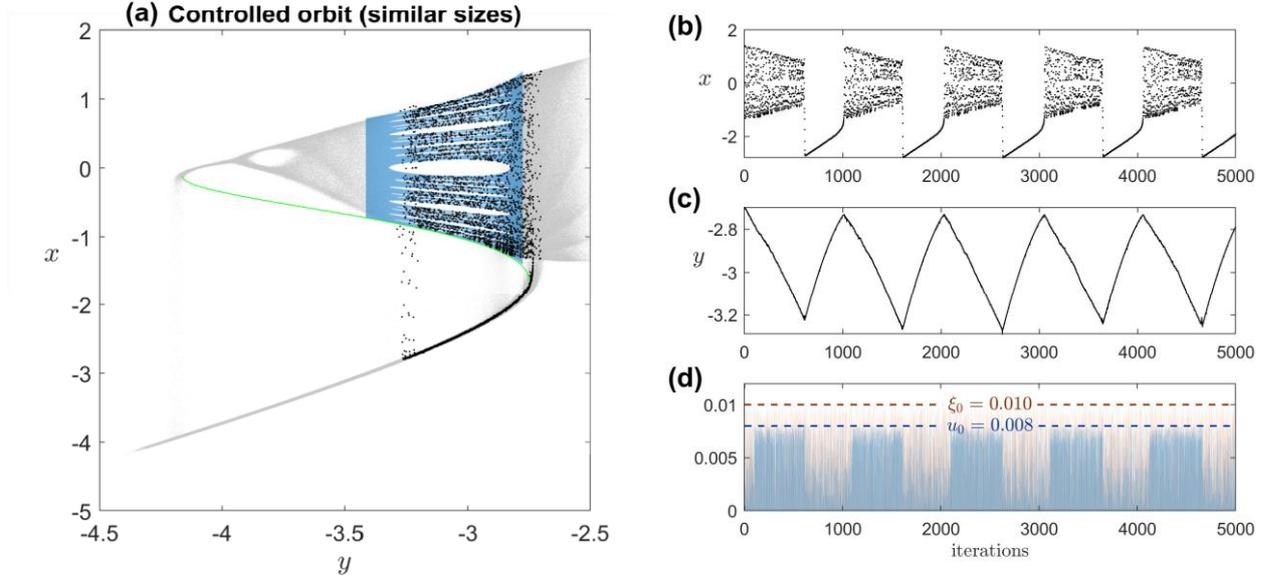

Figure 9: **Control to obtain cycles with similar size.** (a) In blue the set $S$ computed for $\xi_0$ = 0.010 and $u_0$ = 0.008 . In this set, the control is applied to sustain the orbit (black dots) in the safe set until it reaches 600 iterations in the bursting regime. Then the escape of the orbit from the safe set is forced. In this way we can control exactly the duration of the bursting. (b) The $x$-time series of the controlled orbit. (c) The $y$-time series of the controlled orbit. This variable is affected by disturbances but it looks smooth because of the additional control over it. (d) The disturbances $|\xi_n|$ (orange bars) affecting the orbit and the controls $|u_n|$ (blue bars) applied during the 5000 iterations of the orbit. In this scenario, the control is applied in both variables when the orbit is in $Q$, and is only applied in the $y$-variable when the orbit is outside $Q$ to achieve chaotic cycles with similar size. As a consequence, the amount of control applied is bigger than in the two previous scenarios, but it never exceeds the value $u_0$ = 0.008.

## V. SETS $S$ FOR DIFFERENT VALUES OF THE DISTURBANCE $\xi_0$

In the previous section we have shown the application of the control in three different scenarios where we use the upper disturbance bound $\xi_0$ = 0.010 and the upper control bound $u_0$ = 0.008. However, if the values $\xi_0$ and $u_0$ are different, the set $S$ will be different, as shown in Fig. 10. In order to show how this change affects the controlled orbits, we compute again the three scenarios presented before, but for a different disturbance value $\xi_0$ affecting the map. In one case we choose a bigger disturbance $\xi_0$ = 0.020 and in the other one, a smaller disturbance $\xi_0$ = 0.005.

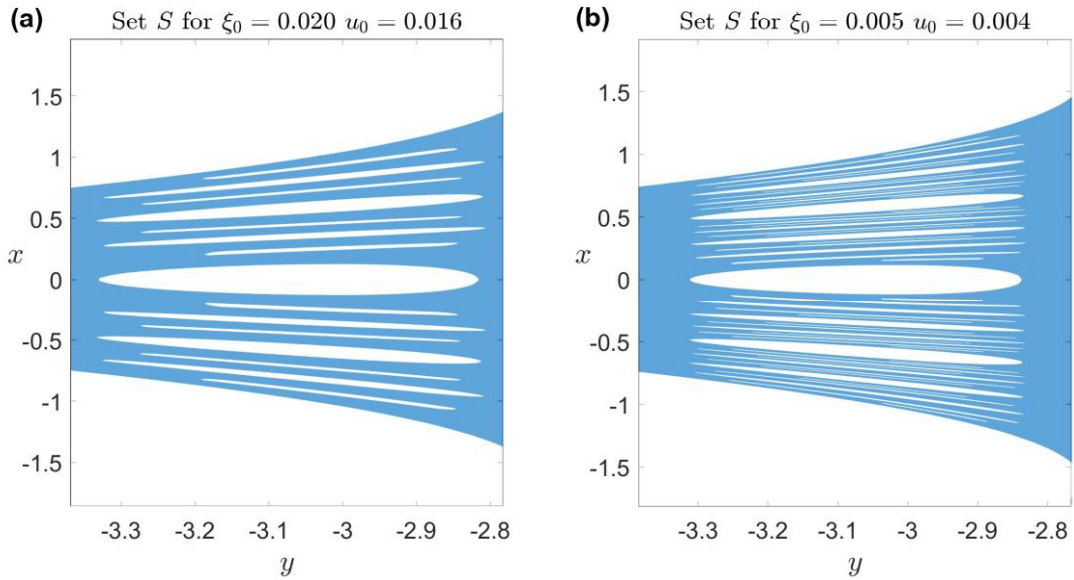

Figure 10: **Computing the set $S$ for different values $\xi_0$.** In both cases the region $Q$ is taken as the rectangle $(y,x) \in [-3.42,-2.78] \times [-1.82,1.92]$. The right and left sides of $Q$ are open boundaries. The grid resolution taken in $Q$ is 2000 × 2000 points. (a) The set $S$ computed for $\xi_0$ = 0.020 and $u_0$ = 0.016. It takes 23 iterations to converges. (b) The set $S$ computed for $\xi_0$ = 0.005 and $u_0$ = 0.004. Note the finer structure for smaller values of $\xi_0$. It takes 37 iterations to converges.

For the case $\xi_0$ = 0.020, we obtain that the minimum upper control bound for which the set $S$ exists is $u_0$ = 0.016 (see Fig. 10(a)) . The corresponding controlled orbits for the three scenarios are shown in Fig. 11.

In the other case, we assume that the upper disturbance bound affecting the map is $\xi_0$ = 0.005. The minimum upper control bound for which the set $S$ exists is $u_0$ = 0.004 (see Fig. 10(b)). The corresponding controlled orbits for the three scenarios are shown in Fig. 12.

These two examples, where different $\xi_0$ have been chosen, reveal the most important feature of the control method. Not only it takes into account the random disturbance affecting the system, but also its intensity, obtaining different sets $S$ that minimize the necessary control in each case.

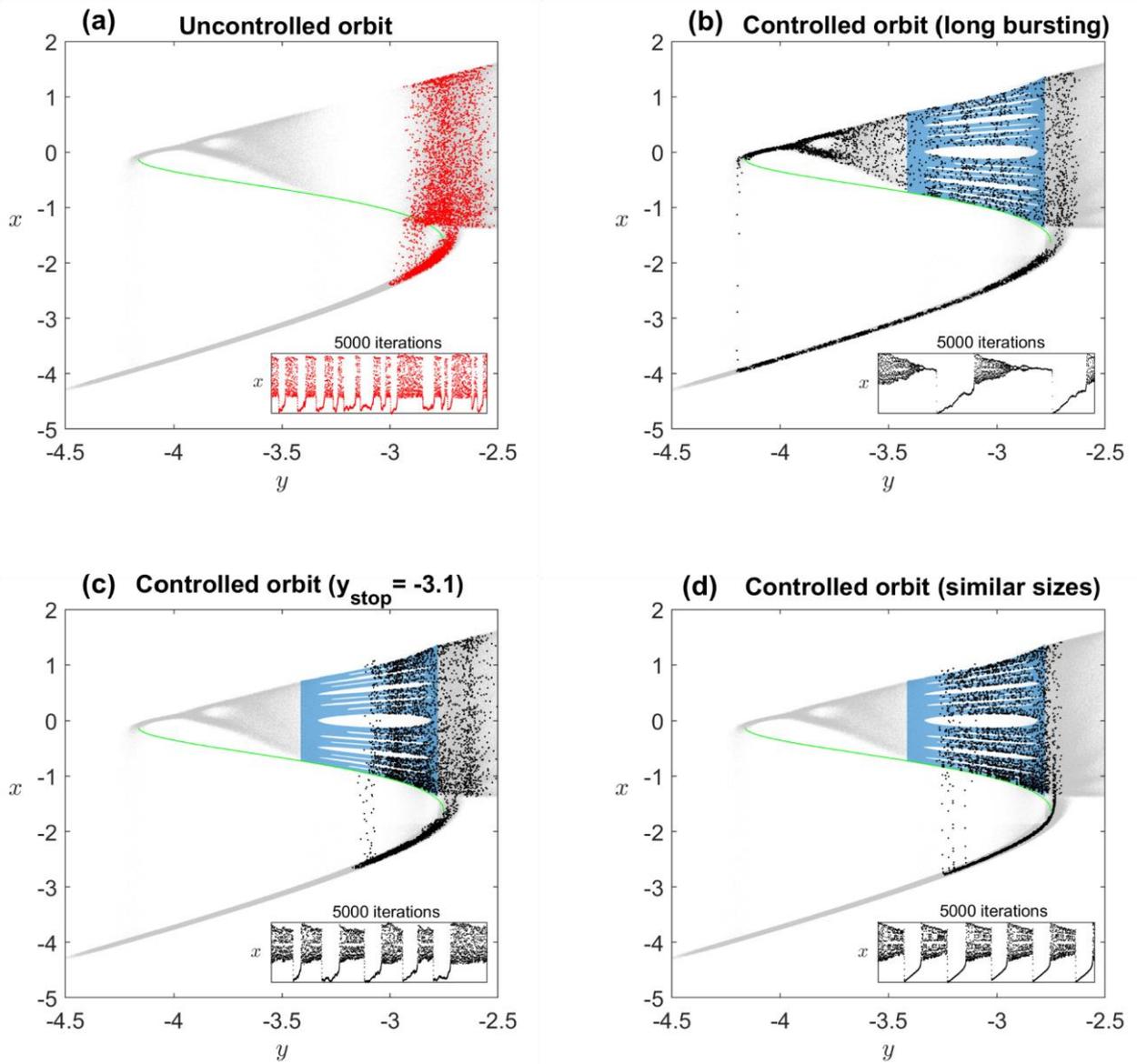

Figure 11: **Big disturbance. Controlling orbits with $\xi_0 = 0.020$ and $u_0 = 0.016$.** Controlled orbits corresponding to the three scenarios presented in the section IV. The only change is the bigger disturbance $\xi_0$ affecting the map and therefore the bigger control $u_0$ required. (a) Uncontrolled orbit. (b) Long bursting size. (c) Control until a specific *y* value in the set *S*. (d) Control to obtain cycles with similar size. The bursting size selected is 600 iterations.

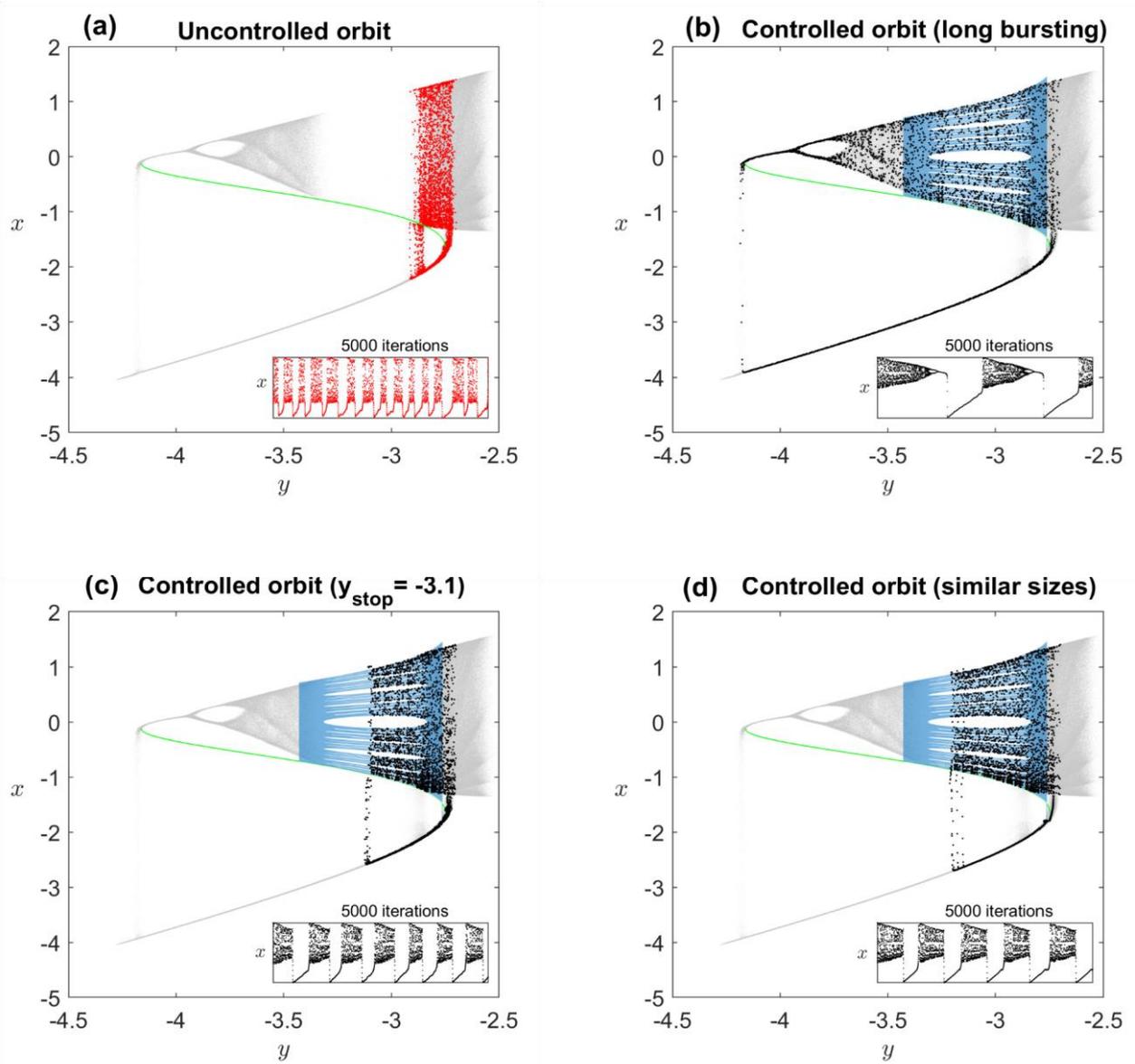

Figure 12: **Small disturbance. Controlling orbits with $\xi_0 = 0.005$ and $u_0 = 0.004$.** Controlled orbits corresponding to the three scenarios presented in the section IV. The only change is the smaller disturbance $\xi_0$ affecting the map and therefore the smaller control $u_0$ required. (a) Uncontrolled orbit. (b) Long bursting size. (c) Control until a specific $y$ value in the set $S$. (d) Control to obtain cycles with similar size. The bursting size selected is 600 iterations.

**VI.     DISTURBANCES AND CONTROL IN ONLY ONE VARIABLE.**

Along this work we have used the following Rulkov map model:

$$x_{n+1} = \frac{\alpha}{(1+x_n^2)} + y_n + \xi_n^x + u_n^x \qquad (5)$$

$$y_{n+1} = y_n - \sigma x_n - \beta + \xi_n^y + u_n^y \quad ,$$

where we consider that both variables were affected by disturbances and both variables can be controlled. However, to complete our study we report here a brief analysis when either one variable is not controlled or is not affected by the disturbances. The results that we obtain, can be summarized in the following three cases.

**Case a)** $\xi_n^x \neq 0$, $u_n^x \neq 0$, $\xi_n^y \neq 0$, $u_n^y = 0$. If we observe the sets $S$ computed before, they are made of approximately horizontal stripes. Typically, the orbit jumps from one stripe to another until it falls outside $S$. In that moment, the control is applied to return the orbit back to the nearest stripe. Due to the horizontal distribution of the stripes, the control applied is mainly in the vertical axis (x-axis). For this reason, if we compute the set $S$ allowing only control in the variable $x$, the set $S$ that we obtain is very similar to the ones computed in the previous sections. The only difference is that the minimum value $u_0^{min}$ for which the set $S$ exists, it is slightly larger than the $u_0^{min}$ obtained when the control is allowed in both variables. For example, in the set $S$ computed in the section III we obtain a $u_0^{min} = 0.008$, while in the case of $u_n^y = 0$, we have obtain $u_0^{min} = 0.0085$.

**Case b)** $\xi_n^x \neq 0$, $u_n^x \neq 0$, $\xi_n^y = 0$, $u_n^y = 0$. The sets $S$ that we obtain in this case are very similar with the sets $S$ shown in this work. The reasons are the same as explained in the previous case, the control $u_n^x$ is active. However there is an important change. Due to the absence of disturbance in the slow-variable $y$, the control scheme proposed in the scenario three to get cycles with similar size is not needed since the $y$-variable behaves smoothly. Even though we know that the disturbance affecting the $x$-variable, will affects the $y$-variable in the next iteration of the map, the influence of this disturbance is very small due to the small coupling value $\sigma = 0.001$ in the equations. As a result, the $y$-variable behaves almost as deterministic and therefore the bursting sizes obtained in both, the scenario two and three, are very similar.

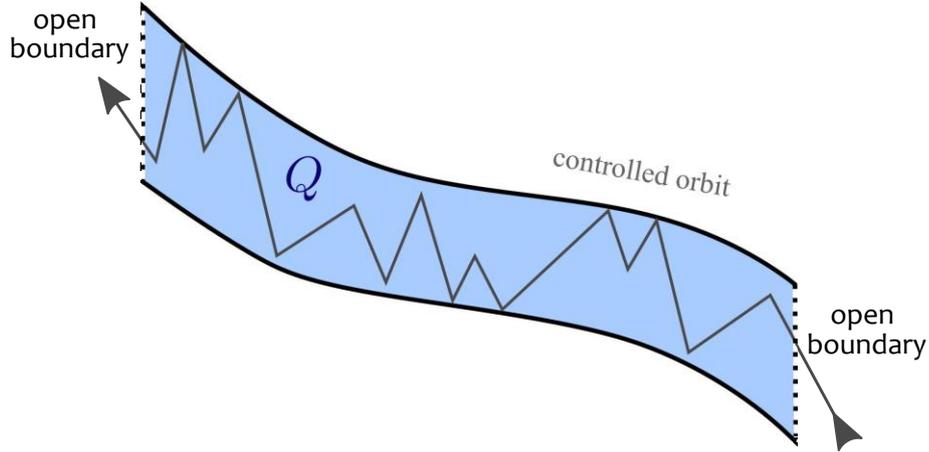

Figure 13: **Schematic control goal.** In blue the region $Q$ defined in the phase space. The open boundaries are indicated with the dashed lines while close boundaries are draw with solid lines. We assume that the dynamics in $Q$ can be modelled as $q_{n+1} = f(q_n) + \xi_n + u_n$ with $|\xi_n| \leq \xi_0$ and $|u_n| \leq u_0$. Given the upper bound of disturbance $\xi_0$ and the upper bound of control $u_0$ the set $S$ (not displayed) can be computed through a recursive algorithm. Orbits in $S$ can be sustained in it applying each iteration of the map a control $|u_n| \leq u_0$, until the orbit escapes (if it escapes) through an open boundary. Black line in $Q$ represents a controlled orbit in $Q$.

**Case c)** $\xi_n^x \neq 0,\ u_n^x = 0,\ \xi_n^y \neq 0,\ u_n^y \neq 0$. If we try to compute sets $S$ controlling only the $y$-variable, we found that it is necessary to apply a very big control resulting in a big $u_0^{min}$ value. Since the $y$-variable is the slow variable, to apply a big control on it will completely destroy the bursting behaviour of the cycles, and for that reason we consider this case (for this map) of no interest, since we want to preserve the chaotic behaviour of the neuron.

## VII. GENERALIZATION OF THE CONTROL METHOD

The control method described in this work has been designed to extend and control the bursting size of a neuron that behaves according to the Rulkov map, Eq. 2. For this case, we define a region $Q$ where the orbits are allowed to enter or abandon it across the right or left boundaries, but not across the top or bottom boundaries. In this way, we were able to extend the bursting size of the neuron.

There might be other systems where the applications of this control scheme can be useful. In general, given a system, we can design a region $Q$ in the phase space, that actuates like a bridge (see Fig. 13) for the orbits to connect regions of the phase space that otherwise would be impossible .

The steps to apply this control technique is summarized as follows:

- Define the region $Q$ in the phase space to connect different regions of phase space. We assume that the dynamics in $Q$ can be described as $q_{n+1} = f(q_n) + \xi_n + u_n$, with $|\xi_n| \leq \xi_0$ and $|u_n| \leq u_0$.

- Define the boundaries behavior (open or close). Orbits are allowed to escape/enter in $Q$ through the open boundaries. Orbits are not allow to escape/enter in $Q$ through the closed boundaries.

- Apply the following recursive algorithm. Beginning with $Q_0 = Q$. The *ith* iteration of the algorithm is:

    1. Fatten the set $Q_i$ by $u_0$ except the open boundaries, obtaining the set denoted by $(Q_i + u_0)$.
    2. Shrink the set $Q_i + u_0$ by $\xi_0$ except the open boundaries, obtaining the set denoted by $(Q_i + u_0 - \xi_0)$.
    3. Let $Q_{i+1}$ be the points $q \in Q_i$, for which $f(q)$ falls inside the set denoted $(Q_i + u_0 - \xi_0)$, or the points $q \in Q_i$ for which $f(q)$ abandon $Q$ through an open boundaries. 4. Return to step 1, unless $Q_{i+1} = Q_i$. We call this final set, the set $S$.

- Control the orbits with the set $S$. Given a point $q \in S$, we evaluate $f(q_n) + \xi_n$ and then we apply the corresponding control $|u_n| \leq u_0$ to put the orbit back in $S$ unless $f(q_n) + \xi_n$ escapes from $Q$ through an open boundary.

Here we want to point out three important considerations. First, this control scheme only describes how an orbit is controlled in the set $S \in Q$. The way the orbit enters in $S$ should be taken into account to design an appropriate region $Q$ in the phase space. For example, in the case of the Rulkov map (see Fig. 4), if we take a bad region $Q'$ as the rectangle $(y,x) \in [-3.5,-3] \times [-1.82,1.92]$ that does not touch the left chaotic region, most of the orbits, after a short chaotic transient, will fall towards the stable manifold at the bottom, and never reaches the right boundary ($y = -3$) of $Q'$. In consequence, very few orbits will enter in $Q'$.

Second, the condition that we establish for the open boundaries (orbits can enter/escape through this boundary), is not well defined since we are working with maps (discrete trajectories) not with flows (continuous trajectories). The criterion that we follow in this work is the simplest one. For a given orbit such that $q_n$ is in $Q$ and $q_{n+1}$ maps outside $Q$, we draw an imaginary straight line between $q_n$ and $q_{n+1}$. If the line crosses the open boundary, we consider that the orbit is abandon $Q$ through the open boundary. If not, we apply the corresponding control $|u_n| \leq u_0$ to put the orbit back in $Q$. This is only one criterion among all the possible choices to define if an orbit crosses an open boundary, and the controller is free to set his own criterion. The steps of the recursive algorithm to obtain $S$ applies in the same way.

Third, this control scheme is designed to be minimally invasive. The control is not applied to guide the orbit from one open boundary to another open boundary. The control scheme is applied to sustain the orbit in $Q$ until, if it happens, the orbit escapes across one of the open boundaries. However, as we show in the subsection IV.C this control technique can be combined with an additional control as long as the controls applied satisfies $|u_n| \leq u_0$.

## VIII. CONCLUSIONS

In this work we propose a control technique to extend the bursting size of a neuron modelled by the two-dimensional Rulkov map affected by bounded disturbances. We assume that the map can be modelled as $q_{n+1} = f(q_n)+\xi_n+u_n$ where the disturbances and controls are bounded so that $|\xi_n| \leq \xi_0$ and $|u_n| \leq u_0$. The control method defines a region $Q$ in the phase space between two separated chaotic regions. To connect both chaotic regions an allow the orbits to exhibits long bursting, we compute an special subset $S \subset Q$ where the orbits can be sustained with minimal control $u_0$. Once the set $S$ is obtained we consider three scenarios of application.

In the first scenario, the control is applied in all the set $S$ to lead the orbit from one chaotic region to the other, resulting in a long bursting size. In the second scenario, we stop the bursting when the orbit reaches a predefined $y$-value in $Q$ resulting in shorter bursting sizes. In the third scenario we stop the burting when it reaches a certain number of iterations. In addition, in this last case, we add an extra control in the $y$-variable to achieve similar cycles with approximately the same bursting size. In all the scenarios, we show how the $S$ adapts for different upper disturbance bound $\xi_0$ to minimize the upper control bound $u_0$ necessary to sustain the orbits in $Q$.

After that, we report the case in which only one variable is controlled showing that the control in the $x$-variable is necessary to keep the chaotic behaviour of the neuron. Finally, we have explained the generalization of the method, in case of its potential application to other systems.

## IX. ACKNOWLEDGMENT

This work has been supported by the Spanish State Research Agency (AEI) and the European Regional Development Fund (ERDF, EU) under Project No. PID2019-105554GB-I00.